\documentclass[12pt]{article}

\usepackage{arxiv}
\usepackage[utf8]{inputenc} 
\usepackage[T1]{fontenc}    
\usepackage{hyperref}       
\usepackage{url}            
\usepackage{booktabs}       
\usepackage{amsmath,amssymb,amsbsy,amsfonts,amsthm}       
\usepackage{nicefrac}       
\usepackage{microtype}      
\usepackage{xcolor}
\usepackage{times}
\usepackage{float}
\usepackage{wrapfig}

\usepackage[sort,compress]{cite} 
\usepackage[margin=0pt,labelsep=space,footnotesize,labelfont=bf,textfont=it,up,belowskip=-8pt,aboveskip=5pt]{caption}
\usepackage{subfigure}
\usepackage{graphics,epsfig,graphicx,color}
\usepackage{algorithm,algorithmic}

\makeatletter

\renewcommand{\paragraph}{\@startsection{paragraph}{4}{\z@}%
 {0.7ex \@plus 0.25ex \@minus .15ex}{-1em}{\normalfont\normalsize\bfseries}} 

\makeatother
\newcommand{\mynote}[3]{
	\textcolor{#2}{\fbox{\bfseries\sffamily\scriptsize#1}}
		{\textsf{\emph{#3}}}
}

\newcommand{\tb}[1]{\mynote{Tan}{magenta}{#1}}

\definecolor{sblue}{cmyk}{0.98,0.13,0,0.43} 
\definecolor{sblue}{cmyk}{0.98,0.13,0,0.43} 

\definecolor{sred}{cmyk}{0.01,0.98,0,0.2} 
\reversemarginpar
\newcommand{\mmargin}[1]{{\marginpar{\em\tiny #1}}}\renewcommand{\mmargin}[1]{}

\newcommand{\mb}[1]{\mathbf{#1}}
\newcommand{\mc}[1]{\mathcal{#1}}

\newcommand{\bs}[1]{\boldsymbol{#1}}

\newcommand{\LRp}[1]{\left( #1 \right)}

\newcommand{\bi}{\begin{itemize}}
\newcommand{\ei}{\end{itemize}}

\newcommand{\tasklistbegin}{
\begin{list}{\textbf{Task \Alph{project}-\arabic{taskctr}}.}{\usecounter{taskctr}}
\setcounter{taskctr}{\value{task}}
}
\newcommand{\tasklistend}{
\setcounter{task}{\value{taskctr}}
\end{list}
}
\newcounter{rtaskno}

\newcommand{\F}{\mb{F}}

\renewcommand{\u}{u}

\newcommand{\ub}{\mb{\u}}

\renewcommand{\mynote}[3]{
	\textcolor{#2}{\fbox{\bfseries\sffamily\scriptsize#1}}
		{\small$\blacktriangleright$\textsf{\emph{#3}}$\blacktriangleleft$}
}

\newcommand{\w}{w}
\newcommand{\wb}{\bs{\w}}

 \newcommand{\db}{\mb{y}}

\newcommand{\reg}[1]{\mc{R}(#1)}
\newcommand{\dbp}{\tilde{\db}(\wb, \ub)}
\newcommand{\dbo}{\db^{\text{obs}}}
\newcommand{\epsn}{\varepsilon_{\text{NN}}}
\newcommand{\BO}{C}

\title{Accelerating PDE-constrained Inverse Solutions with Deep Learning and Reduced Order Models}
\author{
    Sheroze Sheriffdeen \\
    Oden Institute for Computational Engineering and Sciences \\
    The University of Texas at Austin \\
    Austin, TX 78712, USA \\
    \texttt{sheroze@oden.utexas.edu} \\
    \And
    Jean C. Ragusa \\
    Department of Nuclear Engineering \\
    Texas A\&M University \\
    College Station, TX 77843, USA \\
    \texttt{jean.ragusa@tamu.edu} \\
    \And
    Jim E. Morel \\ 
    Department of Nuclear Engineering \\
    Texas A\&M University \\
    College Station, TX 77843, USA \\
    \texttt{morel@tamu.edu}\\
    \And
    Marvin L. Adams \\
    Department of Nuclear Engineering \\
    Texas A\&M University \\
    College Station, TX 77843, USA \\
    \texttt{mladams@tamu.edu}\\
    \And
    Tan Bui-Thanh  \\
    Department of Aerospace Engineering and Engineering Mechanics and\\
    Oden Institute for Computational Engineering and Sciences \\
    The University of Texas at Austin \\
    Austin, TX 78712, USA \\
    \texttt{tanbui@ices.utexas.edu}\\ 
}

\begin{document}
\maketitle

\begin{abstract}
Inverse problems are pervasive mathematical methods in inferring knowledge from  observational and experimental data by leveraging simulations and models.
Unlike direct inference methods, inverse problem approaches typically require many forward model solves usually governed by Partial Differential Equations (PDEs). This a crucial bottleneck in determining the feasibility of such methods. 
While machine learning (ML) methods, such as deep neural networks (DNNs), can be
employed to learn nonlinear forward models, designing a network
architecture that preserves accuracy while generalizing to new parameter regimes is a
daunting task. Furthermore, due to the
computation-expensive nature of forward models, state-of-the-art black-box ML
methods would require an unrealistic amount of work in order to obtain an
accurate surrogate model. On the other hand, standard Reduced-Order
Models (ROMs) accurately capture supposedly important physics
of the forward model in the reduced subspaces, but otherwise could be
inaccurate elsewhere. In this paper, we propose to enlarge the validity of ROMs and
hence improve the accuracy outside the reduced subspaces by incorporating a data-driven ML
technique. In particular, we focus on a goal-oriented approach that substantially improves the accuracy of reduced models by learning the error between the forward model and the ROM outputs. Once an ML-enhanced ROM is constructed it can accelerate the performance of solving many-query problems in parametrized forward and inverse problems. Numerical results for inverse problems governed by elliptic PDEs and parametrized neutron transport equations will be presented to support our approach.
\end{abstract}


\section{Introduction}

Data-driven machine learning techniques provide a useful framework for modeling and control of complex systems of scientific and industrial interest. Examples include physics-informed neural networks\cite{raissi2017physics, raissi2019physics}, data-driven discovery of governing equations\cite{brunton2016discovering}, a deep learning-based numerical algorithm for solving variational problems\cite{weinan2018deep}, and more\cite{nguyen2019deep, milano2002neural, wang2016physics}.

Bridging the gap between numerical simulations and data-driven machine learning techniques is important in addressing the increasing need for better accuracy and faster computational performance when dealing with large-scale complex systems. Model order reduction provides a framework to construct efficient and low-dimensional models to alleviate the computational burden of simulating many parameter realizations of systems by capturing the dominant modes of the physical system\cite{benner2015survey}. Recent efforts provide evidence supporting the combination of machine learning techniques with low-dimensional physical models to perform efficient and accurate modeling of physical systems. Examples include modeling statistical error surrogates of reduced order models\cite{drohmann2015romes}, neural network closure models for projection-based reduced models\cite{san2018extreme, san2018neural}, and deep learning model order reduction methods\cite{hartman2017deep}.

In this paper, we shall combine the effectiveness of
traditionally physics-preserving approaches with deep
learning. To achieve this objective, we propose  first to learn the important/dominant physics of the
    problem under consideration by a well-established approach 
    and then to use deep learning to learn the discrepancy of the former with respect to high-fidelity simulations and observation data. In particular, 
    we  will leverage
  the projection-based reduced-order modeling (ROM) approach\cite{benner2015survey}. 
  The ROM is constructed not only to be
  fast-to-evaluate but also to respect the underlying important
  physics. Clearly, while reduced models are designed to be as
  accurate as possible, their
  accuracy is inversely proportional to the
  dimension of the reduced subspace and the complexity of the problem under
  consideration\cite{benner2015survey}.

    {\em The proposed DL-enhanced ROM approach
    induces several advantages over the existing methods}: 1) compared
  to standard DL techniques, it preserves physical properties
  including conservation and stability (e.g. via constrained
  least-squares ROM approach); 2) compared to conventional ROM
  approaches, ours is corrected by a NN in a data-driven fashion and hence more accurate; 3) it
  is significantly more reliable than the standard DL surrogate methods as its
  accuracy is first obtained through reduced models and then enhanced
  by a NN; and 4) our approach decouples physic-embedding (via reduced
  models) and accuracy (via NN), and thus allowing standard DL methods
  to be employed in a straightforward fashion.

\section{Parameter-to-Observable Maps and Many-Query Problems}
Consider the parameter-to-observable map,
\begin{equation}
\db := \F\LRp{\wb(\ub)},
\end{equation}
where the state $\wb \in \mathbb{R}^N$ depends on the parameters $\ub \in \mathbb{R}^{N_u}$ through a forward model usually represented as a partial differential equation and $\db \in \mathbb{R}^d$ represents the measurements of the quantities of interest. The parameter-to-observable map is evaluated by first solving a dynamical system for the state $\wb$ such that,
\begin{equation}
	\mathbf{R}(\wb, \ub) = 0
\end{equation}
and extracting observations from the state,
\begin{equation}
 \db = \BO \wb(\ub)
 \end{equation}
where $\mathbf{R}$ is some discrete operator, for example a residual operator resulting from a numerical discretization of a set of partial differential equations and $\BO \in \mathbb{R}^{d \times N}$ is the linear operator computes the quantities of interest given the state variables $\wb(\ub)$. 

 \subsection{Model Order Reduction}
 Model order reduction techniques aim to reduce the computational burden of evaluating the parameter-to-observable map by constructing reduced models that are faster and cheaper to evaluate. Projection-based model order reduction techniques employ a reduced-space basis $\Phi$ to form a linear subspace $\mathcal{V}^m \subset \mathbb{R}^N$ defined to be the span of the reduced-basis onto which the governing equations are projected.  An approximate solution $\tilde{\wb}$ is computed in the reduced space:
 \begin{equation}
 	\wb(\ub) \approx \tilde{\wb}(\ub) = \Phi \wb_r(\ub)
 \end{equation}
 where $\wb_r \in \mathbb{R}^m$ is the vector of coordinates of the approximate solution $\tilde{\wb}$ and is the solution to reduced system,
 \begin{equation}
 	\Psi^T \mathbf{R}(\Phi\wb_r, \ub) = 0, \quad \db_r = \BO \Phi\wb_r(\ub)	
 \end{equation}
 where $\Psi$ is the basis for the test space $\mathcal{W}^m \subset \mathbb{R}^N$. To efficiently construct the reduced trial basis, a model-constrained adaptive sampling procedure (see, e.g., \cite{bui2007goal} and the references therein) is used to obtain high-fidelity state solution snapshots by greedily exploring the parameter space in order to find points leading to the largest errors in the quantities of interest.
 
 Although reduced models can provide an accurate estimate of $\db(\wb, \ub)$, the inherent limitation introduced by approximating the state space with a linear subspace can introduce errors $\varepsilon(\ub)$ in the quantity of interest: \[ \db(\wb, \ub) =  \db_r(\wb_r, \ub, \Phi) + \varepsilon(\ub). \]
 
 

\section{Improving Reduced Order Models with Deep Learning}
 
 \subsection{Data-driven Discrepancy Function}

We propose to learn the error term $\varepsilon(\ub)$ of the ROM using a deep learning model, equipped with the capacity to learn the possibly nonlinear dependency of the quantity of interest as a function of its parameters. The DL-enhanced quantity of interest $\tilde{\db}(\ub)$ would incorporate the physics preserved by the ROM with the approximation capacity of a deep neural network\cite{hornik1989multilayer} to provide a computationally efficient parameter-to-observable map. The current work focuses on improving parameter-to-observable maps for time-independent systems. The proposed DL-enhanced ROM quantity of interest reads
\begin{equation}
    \tilde{\db}(\ub) = \db_r \LRp{\ub, \Phi} + \varepsilon_{\mathrm{NN}}(\ub,\bs{\theta})
\end{equation}
where $\Phi$ is the reduced trial basis. 

Learning a corrective term $\varepsilon_{\mathrm{NN}}(\ub,\bs{\theta})$, through finding the "best" network hyper-parameters $\bs{\theta}$, to augment a reduced model provides a mechanism to approximate possibly non-linear dependencies between the observables and the parameters in an efficient manner. Furthermore, the learned neural network model is data-driven, providing the ability to incorporate experimental observation data and high-fidelity simulation data during the offline stage of model creation. The learned model can then be incrementally improved in an online fashion during the prediction phase. 

A purely data-driven surrogate model relies on a rich training dataset to learn the dynamics of the system whereas reduced order models, particularly projection-based reduced models generated with model-constrained adaptive sampling\cite{bui2007goal}, efficiently albeit intrusively learns to approximate the high-fidelity model. Then, the corrective term will be easier to train compared to the purely data-driven surrogate model since the contained knowledge of dominant physics of the system would have already been captured in the ROM and will be transferred to the combined prediction. 


\subsection{Training the Model}
To train the deep learning corrective model, an offline data-driven stage is required. The training set $\{(\ub^i,\db^i-\db^i_r)\}_{i=1}^{N_{\text{t}}}$ is formed by simultaneously running high-fidelity forward model and ROM for the same parameters $\ub_i$. Each quantity of interest $\db^i(\ub^i)$ could even be obtained from experimental data in the case where a feasible high-fidelity model is not present. 

For parameters in infinite-dimensional function spaces, a finite element discretization scheme is used enabling this corrective mechanism to be used in a wide range of physical applications. The deep learning model admits the nodal values of the parameters in the input layer.


\subsection{Hyper-parameter Optimization}
Learning the error between the full model and the reduced model is a high-dimensional deep learning regression problem whose performance is sensitive to the particular architecture of the deep neural network model. The pertinent hyper-parameters dictating the architecture of the model include:
\begin{itemize}
	\item number of hidden layers
	\item number of weights per hidden layer
	\item choice of nonlinear activation function between layers
	\item choice of the optimizer to train the weights and its associated learning rate
	\item batch size for mini-batch gradient descent
	\item number of epochs to perform training
	\item dropout probability\cite{srivastava2014dropout}
	\item presence of residual connections\cite{he2016deep}
	\item weight regularization coefficients
\end{itemize}	

We use a Bayesian optimization framework to pick hyper-parameters for the NN discrepancy function\cite{snoek2012practical}. A Gaussian process\cite{rasmussen2003gaussian} surrogate model is employed to parametrize the performance of the neural network as a function of the hyper-parameters. New hyper-parameters are sampled by minimizing an acquisition function. For this work, scikit-optimize's \texttt{gp\_minimize}\cite{head2018scikit} was used as the optimizer which chooses an acquisition function in a probabilistic manner so as to minimize either the negative the expected improvement, the lower confidence bound of the Gaussian process, or the negative probability of improvement at each sampling step. This approach to pick hyper-parameters improves on grid search and random search as the sampling procedure better explores the hyper-parameter space by leveraging information from previous neural network architectures. 

\section{Parameter Inference}

\subsection{Deterministic Inverse Problems with Reduced Models and Discrepancy Functions}
Given observations of the state variable, a common many-query problem is to infer the relevant parameters that caused the observations. The following PDE-constrained optimization procedure aims to reconstruct the parameters by minimizing the objective function $J(\wb, \ub) : \mathbb{R}^N \times \mathbb{R}^{N_u} \rightarrow \mathbb{R}$ by solving
\begin{equation}
\begin{aligned}
\underset{\ub}{\min} \quad & J(\wb, \ub) := \dfrac{1}{2} ||\dbo - \dbp ||^2 + \gamma \reg{\ub} \\
\text{s. t.} \quad &  g(\wb_r, \ub) = \Psi^T \mathbf{R}(\Phi\wb_r, \ub) = 0
\end{aligned}
\end{equation}
where $\dbo$ represents the measured observations, $\reg{\ub}$ is the regularization operator introduced to make the inverse problem well-posed, $\gamma$ is the coefficient of regularization, and the corrected prediction of the observation $\dbp = \db_r(\ub, \Phi) + \epsn(\ub,\bs{\theta})$ combines the reduced quantity of interest and the learned corrective term.

To minimize the above objective function, the gradient with respect to the parameters is computed and optimized using first-order methods (e.g. using L-BFGS). Each gradient evaluation requires solving an adjoint equation and computing the parameter-to-observable map for a new parameter value by solving the reduced system. The gradient of the objective function with respect to $\ub$ is given by,
\begin{equation}
\nabla_{\ub} J(\ub) = \lambda^T\dfrac{\partial g(\wb_r, \ub)}{\partial \ub} + (\db_{\text{obs}} - \dbp )^T \nabla_{\ub} \epsn(\ub,\bs{\theta}) + \gamma \nabla_{\ub} \reg{\ub}
\end{equation}
where $\lambda \in \mathbb{R}^d$ is the solution to the adjoint equation (see Appendix A),
\begin{equation}
\left(\dfrac{\partial g(\wb_r, \ub)}{\partial \wb_r}\right)^T \lambda = (\BO \Phi)^T(\db_{\text{obs}} - \dbp )
\end{equation}

The gradient of the DL discrepancy function $\nabla_{\ub} \epsn(\ub,\bs{\theta})$ is computed using automatic differentiation of the network using Tensorflow\cite{abadi2016tensorflow} and the partial derivatives $\dfrac{\partial g(\wb_r, \ub)}{\partial \ub}$ and $\dfrac{\partial g(\wb_r, \ub)}{\partial \wb_r}$ are computed using the relevant reduced equations.

\section{Numerical Experiments} 

\subsection{Steady Heat Conduction Problem}

In this numerical experiment, a reduced order model corrected with the deep learning discrepancy function is used to compute quantities of interest as a function of temperature in a steady-state heat conduction problem. The temperature distribution within the fin, $\wb$, is governed by the following elliptic partial differential equation:
\begin{alignat}{3}
    - \ub \nabla^2 \wb &= 0 \quad&&  \mathrm{ in }\quad \Omega \\
    - \ub (\nabla \wb \cdot \hat{\mathbf{n}}) &= \mathrm{Bi}\, \wb \quad && \mathrm{ on }\quad \Gamma^{\mathrm{ext}} \setminus \Gamma^{\mathrm{root}} \label{eq:ext}  \\
    - \ub (\nabla \wb \cdot \hat{\mathbf{n}}) &= -1 \quad &&  \mathrm{ on }\quad  \Gamma^{\mathrm{root}} \label{eq:root}
\end{alignat}
where $\ub$ denotes the thermal heat conductivity, $\mathrm{Bi}$ is the Biot number, $\Omega$ is the physical domain describing the thermal fin, $\Gamma^{\mathrm{root}}$ is the bottom edge of the fin, and $\Gamma^{\mathrm{ext}}$ is the exterior edges of the fin. Equation (\ref{eq:ext}) models convective heat losses to the external surface, and equation (\ref{eq:root}) models the heat source at the root. The parameters of interest are the thermal heat conductivity values of the fin and observations are functions of the temperature $\wb$. 
\begin{figure}[H]
    \centering
    \includegraphics[scale=0.4]{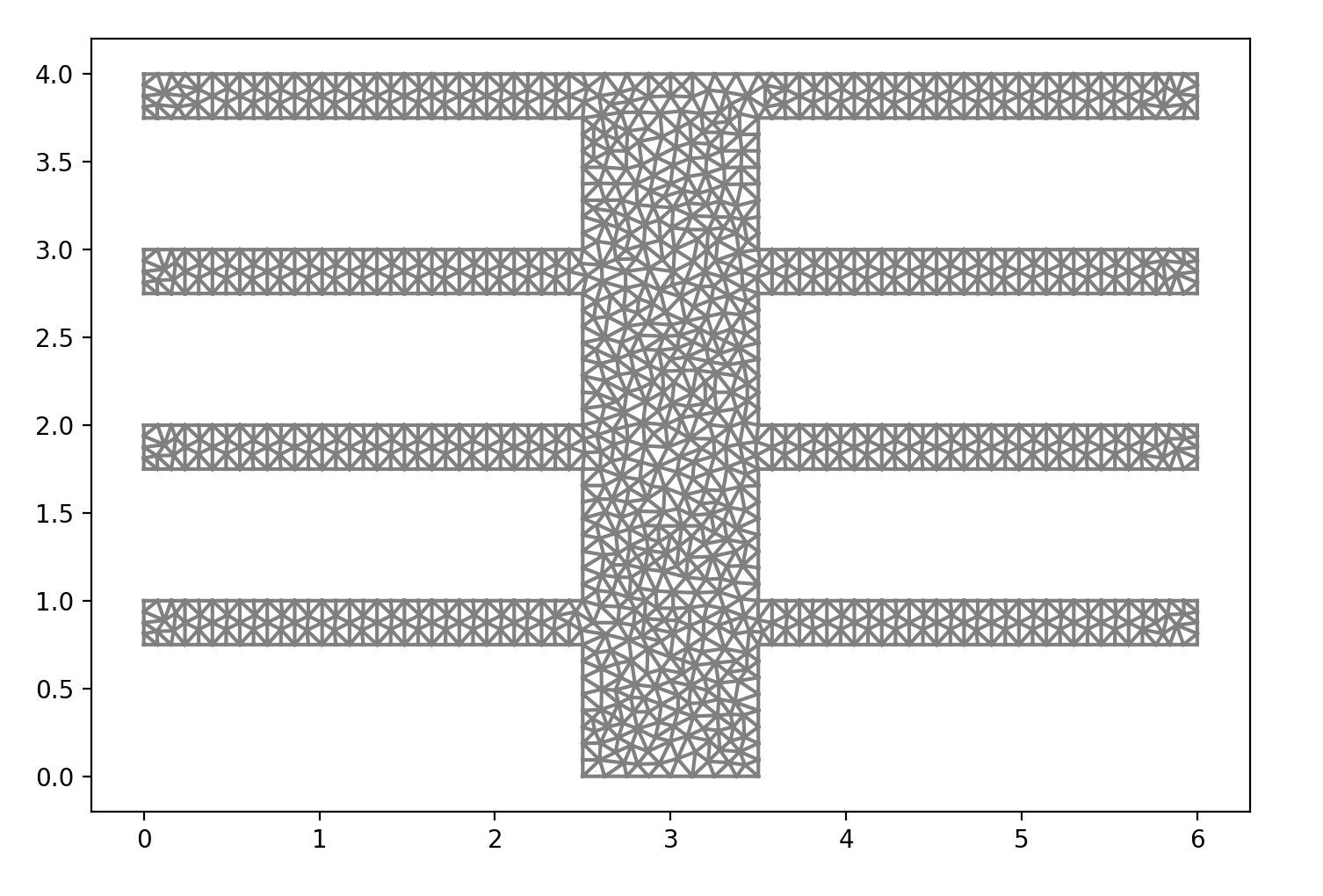}
    \caption{Finite element mesh for the thermal fin with 1446 degrees of freedom}
    \label{fig:my_label}
\end{figure}
The above system of equations can be discretized using the finite element method and the solution can be obtained by solving the linear system,
\begin{equation}
    A(\ub)\wb = B(\ub), \quad \db = C(\ub)\wb
\end{equation}
where $\ub \in \mathbb{R}^{N}$ is the discretized thermal conductivity, $\wb \in \mathbb{R}^{N}$ is the nodal value vector of the temperature distribution, $A(\ub) \in \mathbb{R}^{N \times N}$, $B(\ub) \in \mathbb{R}^{N}$, $\db \in \mathbb{R}^d$ is the quantity of interest, and $C(\ub) \in \mathbb{R}^{d \times N}$ with $N$ being the number of degrees of freedom and $d$ being the number of observables. 

We define the residual for the reduced system given the reduced trial basis $\Phi$ and test basis $\Psi$ as,
\begin{equation}
	R(\Phi \wb_r, \ub) = B(\ub) - A(\ub) \Phi \wb_r
\end{equation}
and the projection-based model order reduction technique yields the reduced system of the form
\begin{equation}
    A_r(\ub)\wb_r = B_r(\ub), \quad \db_r = C_r(\ub)\wb_r
\end{equation}
where $A_r(\ub) = \Psi^T A(\ub) \Phi$, $B_r(\ub) = \Psi^T B(\ub)$, $C_r(\ub) = C(\ub) \Phi$, and $\db_r$ is the reduced quantity of interest. 

The existence of an affine decomposition of the matrix $A$ can further improve computational complexity of the reduced model.
\begin{equation}
A(\ub) = \sum_{i=1}^{q} \sigma_i(\ub) A_i
\end{equation}
where each $A_i$ does not depend on the parameters $\ub$ and $\sigma_i$ is a scalar function of $\ub$ such that $\mathcal{U} \ni \ub \mapsto \sigma(\ub) \in \mathbb{R}^q$.  The form of the reduced matrix then becomes,
\begin{equation}
A_r(\ub) = \sum_{s=1}^{q} \sum_{t=1}^{q} \sigma_s(\ub) \sigma_t(\ub) \Phi^T A_s^T A_t \Phi
\end{equation}
yielding a further improvement in computational efficiency by enabling the precomputation of the matrices $\Phi^T A_s^T A_t \Phi$. 

For the thermal fin heat conduction problem, we average thermal conductivity over each sub-fin to compress the parameter space in order to obtain an efficient approximate affine decomposition of $A$, relying on the capability of the deep learning error model to correct the introduced approximation error. This approximation further accelerates the construction of the reduced order model as the model-constrained optimization problem to identify the reduced basis now involves a search over the $q$-dimensional compressed parameter space as opposed to the much larger $N$-dimensional space. 

The corrected prediction of the quantity of interest thus becomes,
\begin{equation}
    \db(\wb, \ub) \approx \tilde{\db}(\wb, \ub) = \db_r(\wb_r, \sigma(\ub)) + \epsn(\ub,\bs{\theta})
\end{equation}
The reduced space is spanned by an 80-dimensional basis computed by an adaptive model-constrained optimization procedure\cite{bui2007goal}. The deep neural network is trained by obtaining a simulation dataset comprising of parameters $\ub_i$ and their corresponding reduced model errors $\varepsilon_i(\ub) =  \db(\wb, \ub) - \db_r(\wb_r, \sigma(\ub), \Phi)$ by simultaneously solving the high-fidelity model and the reduced model. 

\subsection{Inferring Thermal Conductivity Parameters with Sparse Temperature Observations}
Given observations $\dbo$ of the temperature distribution, the aforementioned deterministic inverse problem is posed in order to infer the thermal conductivity parameters:
\begin{equation}
\min_{\ub} J(\ub) := \dfrac{1}{2} ||\tilde{\db}(\ub) - \dbo ||^2 + \gamma \reg{\ub}
\end{equation}

To evaluate the accuracy of the corrected reduced order model, two different quantities of interest are examined each verified with two numerical experiments. The first two numerical experiments use average temperature measurements in each sub-fin as the quantity of interest $\dbo \in \mathbb{R}^9$. The last two numerical experiments use random point observations on the surface of the thermal fin. The number of surface observations are limited to 40 and the same finite-element discretization with 1446 degrees of freedom is used in all of the experiments. 
For each set of numerical experiments, two forms of the true thermal conductivity are used to generate observations from the high-fidelity model. For the first case, the true conductivity is constant on each sub-fin and total variation regularization is used for the deterministic inverse problem. For the second case, the true conductivity was drawn from a Gaussian random field and Tikhonov regularization was used.

The estimate of thermal conductivity was obtained by minimizing the objective function with a bound constrained limited-memory BFGS routine\cite{byrd1995limited, zhu1997algorithm}. The starting point for the optimization algorithm is drawn from a Gaussian random field and the bounds for the optimization is chosen to be within 5\% of the maximum and the minimum values of the true solution in order to contrast the inversion results to the case where the high-fidelity model is used as the parameter-to-observable map.

\subsubsection{Experiment 1}
The true conductivity is constant on each sub-fin and the observations are averaged temperature per sub-fin. A dense feed forward neural network employed to learn the discrepancy model optimized using Adam\cite{kingma2014adam}. Exponential linear units were used as activation functions in each hidden layer\cite{clevert2015fast}. A \textit{residual block} consisting of two fully-connected layers with skip-connections\cite{he2016deep} followed by batch normalization\cite{ioffe2015batch} was used as the primary building blocks of the deep neural network. The hyper-parameters dictating the architecture of the network were determined by the aforementioned Gaussian optimization procedure to use the Adam optimizer with a learning rate of 3e-4 with scheduled cooling of 50\% every 500 epochs, with a batch size of 400, comprising of three residual blocks with 50 neurons per dense layer, and activated by exponential linear units. The total number of trainable parameters amounted to 80,859.  The trained network has an average validation error of 0.7\% over simulated data generated from the high-fidelity model and the parameters were drawn from a Gaussian random field. 

Table \ref{tb:exp1} shows relative reconstruction errors of the thermal conductivity compared to the true value. The inversion method using the high-fidelity forward model is the most accurate as expected. The DL-enhanced ROM performs with similar accuracy and number of optimizer iterations while the projection-based ROM performs the worst. However, during each optimizer iteration, the high-fidelity model solves an $N \times N$ system (with $N=1446$) whereas the DL-enhanced ROM solves a much smaller $m \times m$ (with $m=80$) system along with an efficient forward and backward pass through the DL discrepancy function with two orders of magnitude smaller number of parameters. 
 
\begin{table}[H]
	\centering
\begin{tabular}{|c|c|c|} \hline
    Method & Relative Reconstruction Error & L-BFGS-B Iterations  \\ \hline
    High-fidelity model ($\gamma = 1e-6$) &  0.1193\% & 468 \\
    Reduced-order model ($\gamma = 1e-6$) & 1.4576\% & 139 \\
    ROM + error correction ($\gamma = 1.5e-5$) & 0.4721\% & 434 \\\hline
\end{tabular}
	\caption{Relative reconstruction error solving the inverse problem where the true solution is piece-wise constant and the observations are average temperatures on each sub-fin}
	\label{tb:exp1}
\end{table}

\begin{figure}[H]
    \centering
    \includegraphics[width=\textwidth]{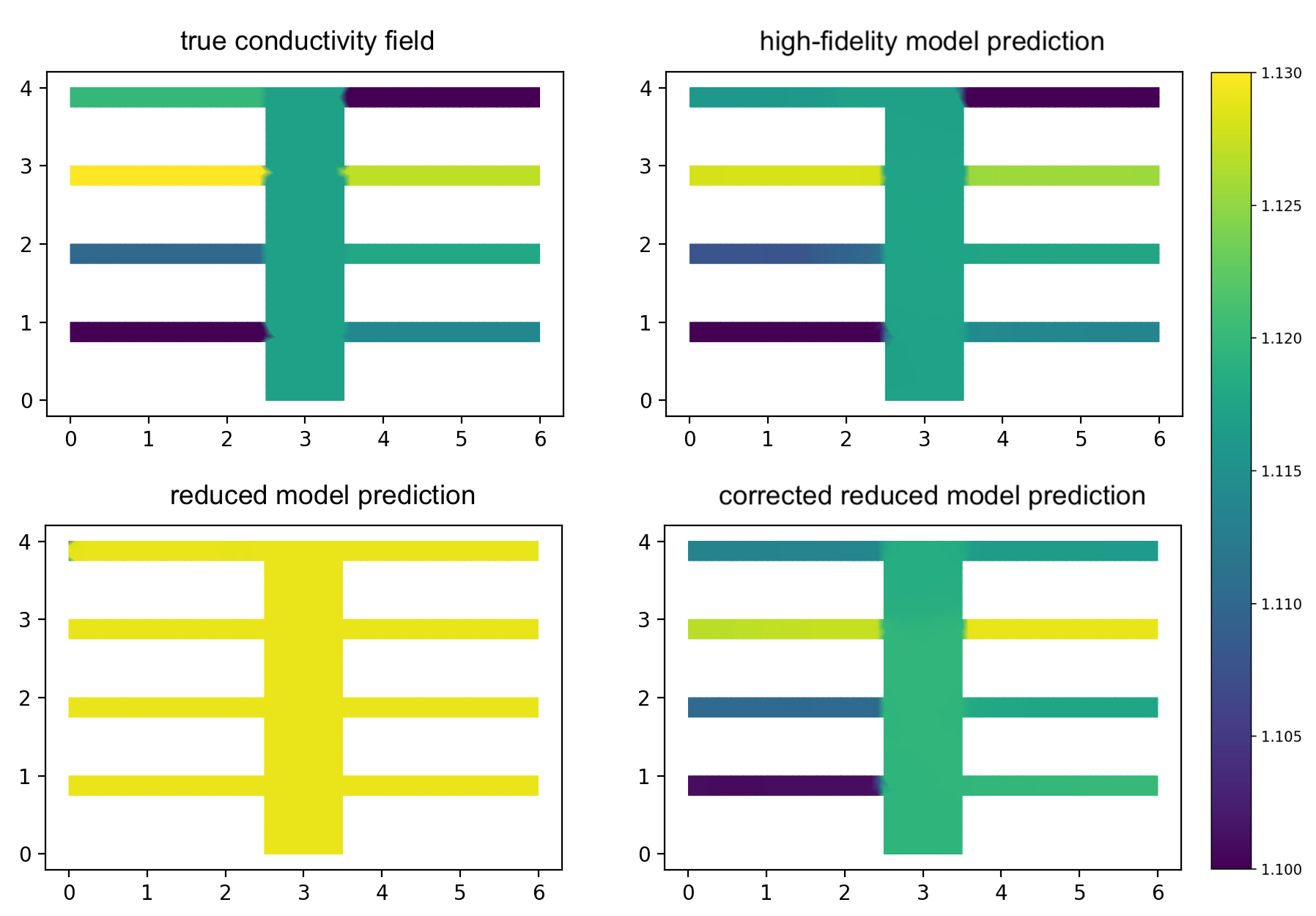}
    \caption{conductivity field reconstructed by solving the deterministic inverse problem compared to the true piece-wise constant conductivity}
\end{figure}

\subsubsection{Experiment 2}
The true conductivity for this experiment is drawn from a Gaussian random field and average temperatures of sub-fins are observed. The discrepancy function trained for the previous experiment was reused.

Table \ref{tb:exp2} shows relative reconstruction errors of the thermal conductivity compared to the true value. Due to the affine approximation of the ROM and the spatially varying nature of the true conductivity field, the relative reconstruction error is significantly larger than the inversion result with the high-fidelity forward model. The DL-enhanced ROM however is able to effectively correct the quantity of interest output leading to similar accuracy compared to the high-fidelity model in a parsimonious fashion.

\begin{table}[H]
	\centering
\begin{tabular}{|c|c|c|} \hline
    Method & Relative Reconstruction Error & L-BFGS-B Iterations  \\ \hline
    High-fidelity model ($\gamma = 1e-6$) &  5.58\% & 609 \\
    Reduced-order model ($\gamma = 1e-6$) & 77.41\% & 150 \\
    ROM + error correction ($\gamma = 1e-6$) & 5.6676\% & 1049 \\ \hline
\end{tabular}
\caption{Relative reconstruction error solving the inverse problem where the true solution is a Gaussian random field and the observations are average temperatures on each sub-fin}
\label{tb:exp2}
\end{table}

\begin{figure}[H]
    \centering
    \includegraphics[width=\textwidth]{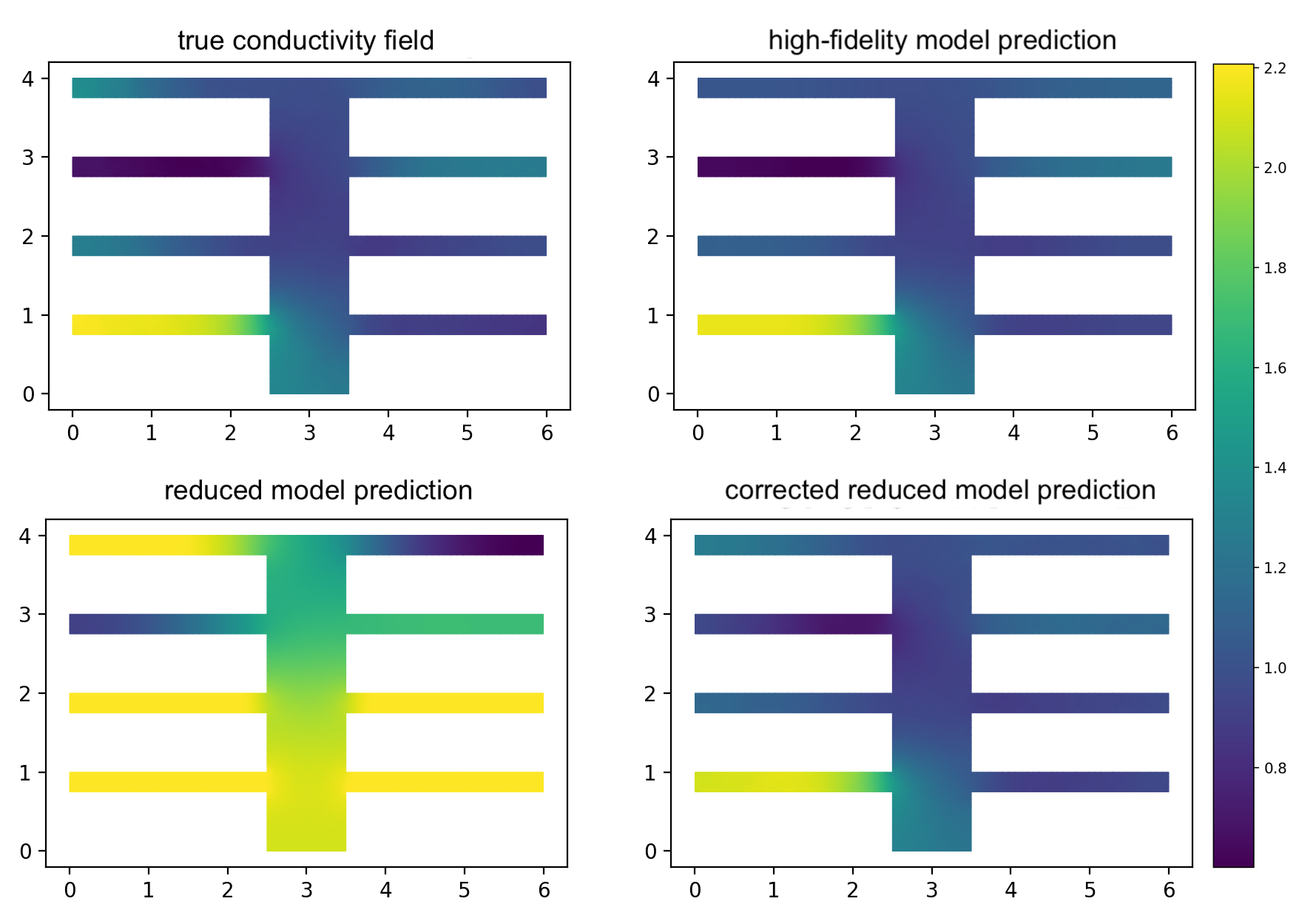}
    \caption{conductivity field reconstructed by solving the deterministic inverse problem compared to the true solution drawn from a Gaussian random field}
\end{figure}

\subsubsection{Experiment 3}
The true conductivity is constant on each sub-fin and random point observations are made on the external surface of the thermal fin. The hyper-parameters dictating the architecture of the network were determined by the aforementioned Gaussian optimization procedure to use the Adam optimizer with a learning rate of 1e-4 with scheduled cooling of 50\% every 500 epochs, with a batch size of 100, comprising of three residual blocks with 100 neurons per dense layer, and activated by exponential linear units. The total number of trainable parameters amounted to 179,840. The trained network has an average validation error of 15\% with simulated data where parameters were drawn from a Gaussian random field. Table \ref{tb:exp3} shows the similar results as experiment 1 with the somewhat worse accuracy of the DL-enhanced inversion attributing to the increase in the number of observations corresponding to higher regression parameters. The number of neural network parameters are still an order of magnitude smaller than the high-fidelity system.

\begin{table}[H]
	\centering
\begin{tabular}{|c|c|c|} \hline
    Method & Relative Reconstruction Error & L-BFGS-B Iterations  \\ \hline
    High-fidelity model ($\gamma = 2e-5$) &  0.1680\% & 747 \\
    Reduced-order model ($\gamma = 1e-6$) & 1.6098\% & 265 \\
    ROM + error correction ($\gamma = 2e-4$) & 0.4962\% & 521 \\\hline
\end{tabular}
	\caption{Relative reconstruction error solving the inverse problem where the true solution is piece-wise constant and the observations are random point observations on the external surface}
	\label{tb:exp3}
\end{table}

\begin{figure}[H]
    \centering
    \includegraphics[width=\textwidth]{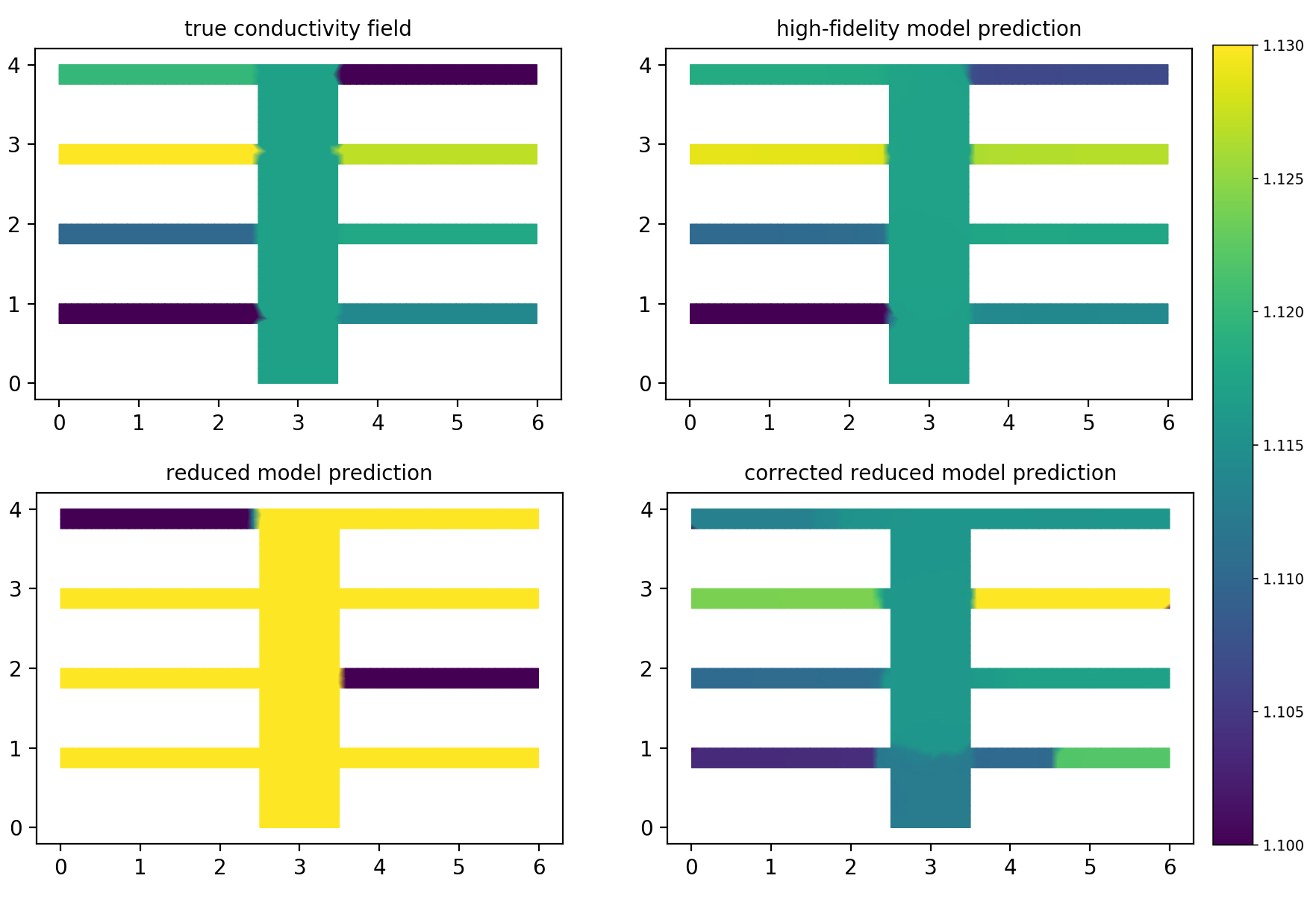}
    \caption{conductivity field reconstructed by solving the deterministic inverse problem compared to the true solution drawn from a Gaussian random field}
\end{figure}

\subsubsection{Experiment 4}
The true conductivity is drawn from a Gaussian random field and random point observations are made on the external surface of the thermal fin. The trained discrepancy function from experiment 3 was reused. Table \ref{tb:exp4} shows results similar to that of experiment 2 with the ROM inversion being significantly worse than the high-fidelity inversion due to the spatially varying nature of the true solution. The enhancing the ROM with the DL discrepancy shows relative reconstruction errors an order of magnitude smaller, although the number of optimizer iterations increase by 80\%. 

\begin{table}[H]
	\centering
\begin{tabular}{|c|c|c|} \hline
    Method & Relative Reconstruction Error & L-BFGS-B Iterations  \\ \hline
    Full-order model ($\gamma = 1e-6$) &  4.4112\% & 724 \\
    Reduced-order model ($\gamma = 1e-6$) & 64.5398\% & 242 \\
    ROM + error correction ($\gamma = 1e-6$) & 7.3075\% & 1321 \\\hline
\end{tabular}
	\caption{Relative reconstruction error solving the inverse problem where the true solution is a Gaussian random field and the observations are random point observations on the external surface}
	\label{tb:exp4}
\end{table}

\begin{figure}[H]
    \centering
    \includegraphics[width=\textwidth]{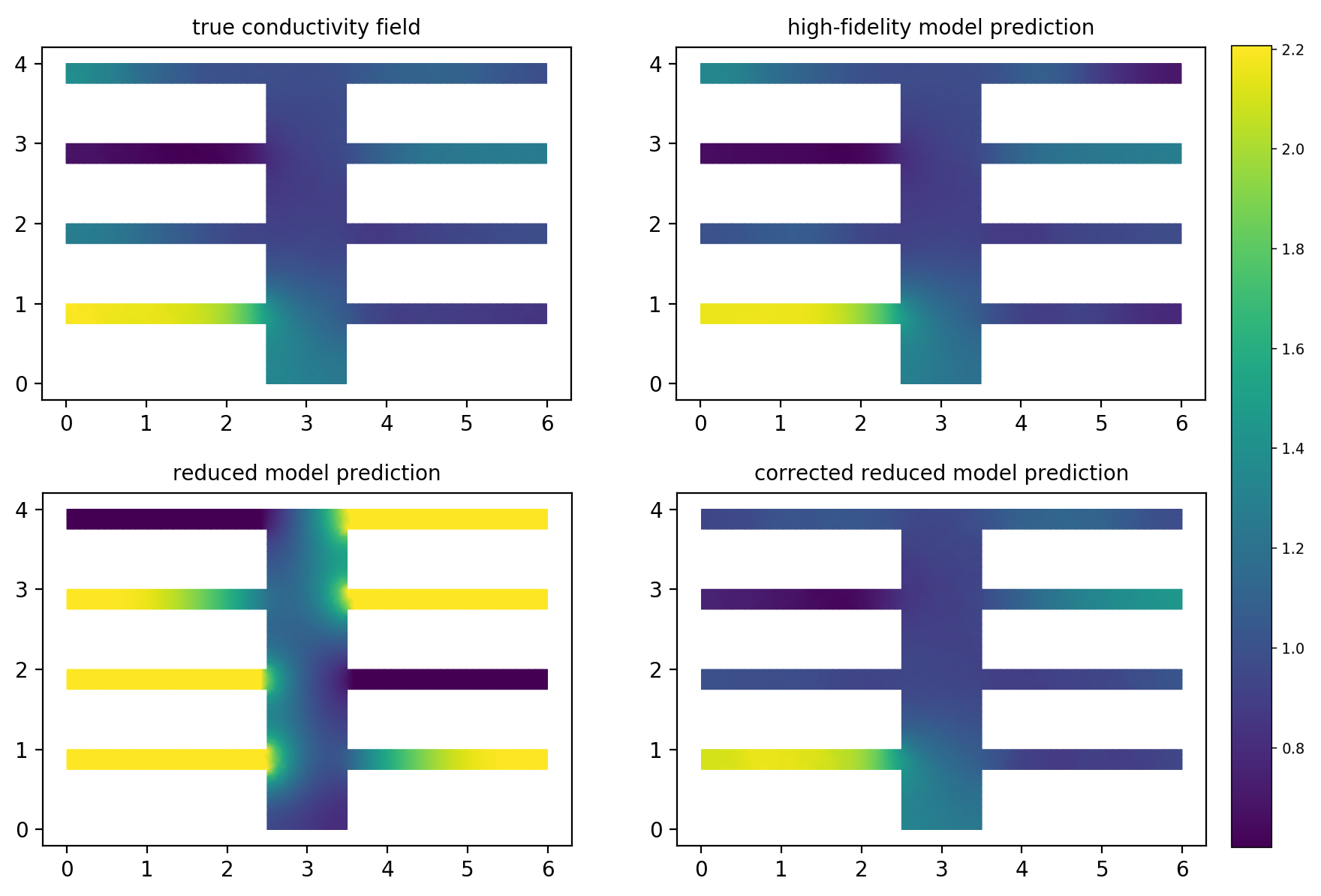}
    \caption{conductivity field reconstructed by solving the deterministic inverse problem compared to the true solution drawn from a Gaussian random field}
\end{figure}

\newpage
\subsection{Physics-Informed Reduced Models for Neutron Transport Equations}

High-fidelity radiation transport calculations require the solution of the Boltzmann equation and are computationally expensive. The DL-enhanced ROM technique is applied to efficiently solve neutral-particle transport modeled by the following equation:

\begin{multline}\label{eq:transport}
\left( \vec{\Omega} \cdot \nabla + \sigma_{t}(\vec{r},E) \right) \Psi^{p}(\vec{r},\vec{\Omega},E) = \\
\sum_{q} \int_{4\pi} d\Omega' \int dE' \,   \sigma^{q \rightarrow p}_{s}(\vec{r},\vec{\Omega} \cdot \vec{\Omega}',E' \rightarrow E) \Psi^{q}(\vec{r},\vec{\Omega}',E') \\
\qquad \qquad \qquad \qquad \qquad \qquad \qquad \qquad + S^{p}(\vec{r},\vec{\Omega},E)
\end{multline}
where $ \Psi^{p}(\vec{r},\vec{\Omega},E)$ is the angular "flux" or "intensity" for particle type $p$, $\vec{\Omega}$ is the unit vector (solid angle) in direction of motion, $\vec{r}$ is the position vector, and $E$ represents its energy. $\sigma_t$ is the macroscopic total interaction cross section, $\sigma_s$ is the scattering cross section, and $S^p$ is the source term. Coupling between particles of different types are resolved iteratively. The phase-space is 6-dimensional with 3 space dimensions, 1 energy dimension, and 2-dimensions for angle. This leads to tens of billions of parameters even with moderate resolution. In order to compute quantities of interest, it is often sufficient to compute the angular integral of the angular flux called scalar flux.
\begin{equation}
    \Phi(\vec{r},E) = \int_{4\pi} d\Omega'  \,  \Psi(\vec{r},\vec{\Omega}',E)
\end{equation}

Physics-informed reduced order models approximate the high-fidelity transport equations in 6-dimensional phase-space with lower-order transport equations in 4-dimensional phase-space. Further reduction in dimensionality with projection-based reduced order models is used on the lower-order physics models as they tend to be large-scale systems. Specifically, a diffusion approximation to the collided component of the total transport flux is used. This approximation is further simplified with energy group collapsing to form discrete energy bins from continuous energy. The energy-dependent diffusion approximation is as follows:
  \begin{equation}
      - \nabla \cdot D_g \nabla \Phi_g + \Sigma_{r,g} \Phi_g = 
      \sum_{g' \neq g} \Sigma_{s,g' \rightarrow g} \Phi_{g'} + Q_g(r), \quad \forall g \in [1,G], \quad \forall r \in \mathcal{D} 
  \end{equation}
with symmetry planes:
  \begin{equation}
      \nabla \Phi_g \cdot n = 0, \forall r \in \partial \mathcal{D}_R
  \end{equation}
and vacuum boundary conditions:
  \begin{equation}
      \Phi_g + \dfrac{D_g}{\alpha} \nabla \Phi_g \cdot n = 0 \forall r \in \partial \mathcal{D}_V
  \end{equation}
 where $g$ is the group index, $G$ the total number of energy groups, $\Sigma_{r,g}$ is the the removal cross section, $D_g$ is the diffusion coefficient, $\Phi_g$ is the flux in group $g$, $\Sigma_{s,g' \rightarrow g}$ is the scattering cross section from group $g'$ to group $g$, and $Q_g$ is the external source of particles in group $g$.
The quantities of interest for this problem are typically functionals of the computed flux solution, for example:
\begin{equation}
      \db = \int_0^{\infty} dE \, \int_{\text{RoI}} d^3r \, \varrho(\vec{r},E) \int_{4\pi} d\Omega \, \Psi(\vec{r}, E, \vec{\Omega}) 
\end{equation}
The quantities of interest can include  dose, dose rates, fluence, fluence rates, and radiation fluxes through boundaries. These quantities of interest can be defined on a subset of the computational domain known as the region of interest and can be transferred to other physics models to compute important biological and electronic effects. The uncertainties in the quantities of interest can arise from factors such as source position, source spectrum, air humidity, ground composition, and location and orientation of the region of interest.
The following results capture the behavior of the projection-based ROM applied to the diffusion approximation compared to the high-fidelity transport solution for the iron-water benchmark. The iron-water benchmark is a standard 1-group 2D benchmark for transport solution techniques comprising of three spatial zones. The sphericized version has been devised and employed to verify our method. The region of interest is the third zone $4 \le x \le 10$.  The diffusion ROM approaches the high-fidelity diffusion approximation with 4-dimensional reduced subspace and is then limited in accuracy compared to the high-fidelity transport solution.

\begin{figure}[H]
    \centering
    \subfigure[Diffusion-based ROM solution flux compared to the transport FOM solution flux. The reduced order model was limited to 20 basis vectors.]{
        \includegraphics[width=0.42\textwidth]{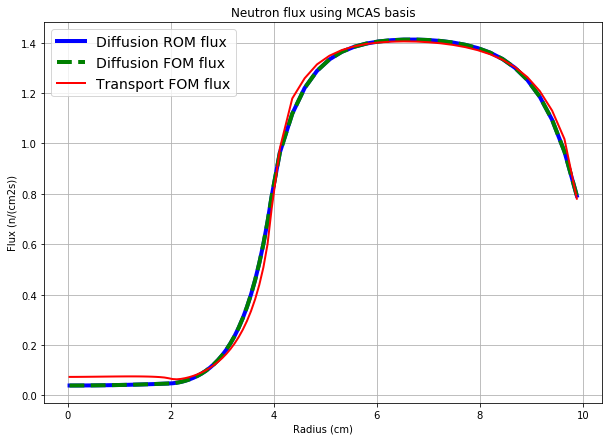}
    }%
    \hfill
    \subfigure[Increasing the dimensionality of the ROM improves relative quantity of interest error compared to the transport solution. The diffusion-based ROM error is limited by the full-order diffusion solution (dotted red line)]{
        \includegraphics[width=0.48\textwidth]{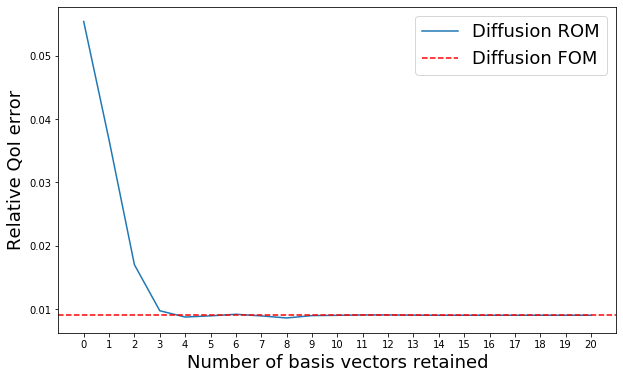}
    }
\end{figure}

The following results incorporate the deep learning discrepancy function to approach accuracy comparable to the high-fidelity transport solution, outperforming the high-fidelity diffusion approximation. The deep learning model is trained using a dataset obtained by simultaneously solving the high-fidelity transport equations and the projection-based diffusion ROM for random removal and scattering cross section values drawn uniformly from the following ranges in each of the three zones:

\begin{table}[H]
	\centering
	\begin{tabular}{| c || c | c |} \hline
	zone & $\Sigma_{r}$ & $\Sigma_{s}$ \\ \hline
	1 & [0.0, 0.2] & [0.7, 1.1] \\
	2 & [0.2, 0.8] & [0.5, 1.8] \\
	3 & [0.4, 0.8] & [0.1, 1.5] \\ \hline
	\end{tabular}
	\caption{Parameter ranges for $\ub = [\Sigma_r, \Sigma_s]$ for the iron-water benchmark}
\end{table}

\begin{figure}[H]
    \centering
    \subfigure[Improvement using the error emulator over a 3-dimensional reduced basis]{
        \includegraphics[width=0.45\textwidth]{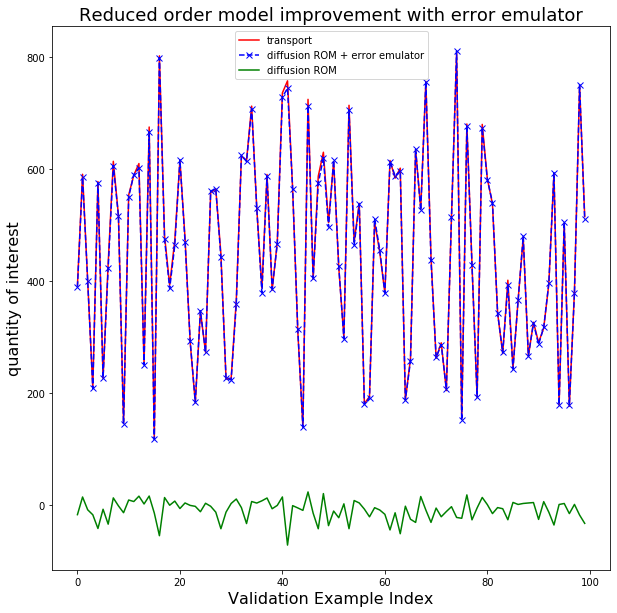}
        \label{fig:dim_three}
    }%
    \hfill
    \subfigure[Improvement using the error emulator over a 22-dimensional reduced basis]{
        \includegraphics[width=0.45\textwidth]{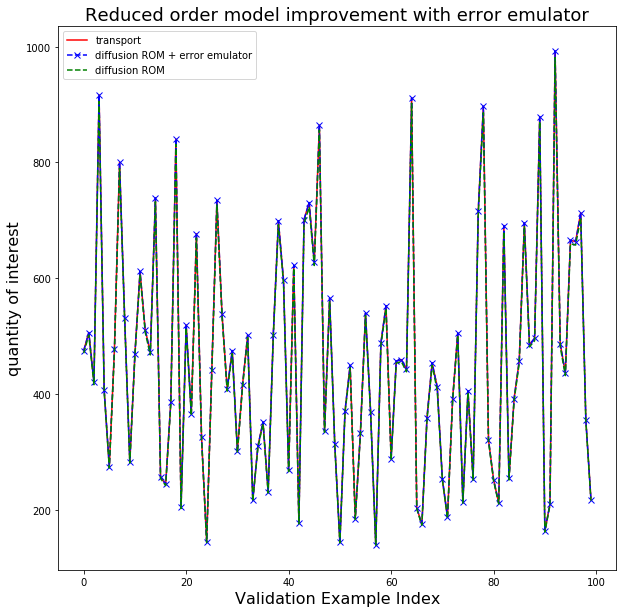}
        \label{fig:dim_twotwo}
    }
    \caption{Parameter-to-observable evaluations for the validation set}
\end{figure}

Figures \ref{fig:dim_three} and \ref{fig:dim_twotwo} show the parameter-to-observable evaluations for a validation set. Each validation set parameter example is used to evaluate the quantity of interest using the high-fidelity transport solution, the diffusion ROM, and the the DL-enhanced diffusion ROM. The discrepancy function is shown to improve even upon the constrained 3-dimensional diffusion ROM.

Table \ref{tb:iw_imp} shows relative quantity of interest error for a validation set in parameter-to-observable maps constructed using the high-fidelity diffusion approximation, diffusion ROMs, and DL-enhanced diffusion ROMs. The ROM variants include a limited 3-dimensional reduced basis and a 22-dimension basis discovered by the model-constrained adaptive sampling. The DL-enhanced ROM variants outperform the high-fidelity diffusion solution showcasing the ability of the discrepancy function to accurately model the ROM error compared to the high-fidelity transport solutions.
    \begin{table}[H]
    \centering
    \begin{tabular}{|l|l|l|} \hline
      Solver Method & QoI error  & QoI error (relative) \\ \hline
      Diffusion (FOM)                         				& 4.252  		& 1.06\%  \\
      Diffusion ROM (3-dim)                         			& 456.228 	& 114.05\% \\ 
      Diffusion ROM (22-dim) 					& 4.345		& 1.08\% \\
      Diffusion ROM (3-dim) + discrepancy function      & 3.464		& 0.86\% \\
      Diffusion ROM (22-dim) + discrepancy function 	 & 0.041		& 0.010\% \\ \hline
    \end{tabular}
    \caption{Average quantity of interest errors compared to the transport solution in the validation dataset}
    \label{tb:iw_imp}
  \end{table}
  
\section{Concluding Remarks}
In this work, we presented a data-driven technique to augment the accuracy of reduced order models by learning their error compared to high-fidelity models and experimental data with the goal of accelerating many-query problems in deterministic inverse problems. We presented preliminary results that support our approach in accelerating parameter-to-observable maps for an elliptic PDE and a parametrized neutron transport problem. We solved a deterministic inverse problem using a DL-enhanced reduced model to efficiently reconstruct thermal conductivity parameters in a steady heat conduction problem given sparse temperature observations.

Future work involves (1) incorporating parameter sensitivities to solve a deterministic inverse problem to infer scattering and absorption coefficients in neutron transport equations, (2) utilizing the DL-enhanced reduced order models to perform Bayesian inference, and (3) exploring physics-informed regularization mechanisms to accelerate the training of the DL discrepancy function.

\newpage

\bibliographystyle{unsrt}
\bibliography{references}

\appendix
\section{Gradient of the Objective Function}
The gradient of the objective function
\begin{equation}
J(\wb, \ub) := \dfrac{1}{2} ||\dbo - \dbp ||^2 + \gamma \reg{\ub}
\end{equation}
with respect to $\ub$ is given by,
\begin{equation}
\nabla_{\ub} J(\wb, \ub) = (\dbo - \dbp) \nabla_{\ub} \dbp + \gamma \nabla_{\ub} \reg{\ub}
\label{eq:J_grad}
\end{equation}
The gradient of the corrected prediction of the observable is given by, 
\begin{equation}
	\nabla_{\ub} \dbp = \dfrac{\partial \db_r \LRp{\ub, \Phi}}{\partial \wb_r} \nabla_{\ub} \wb_r +  \nabla_{\ub} \epsn(\ub)
	\label{eq:pred_grad}
\end{equation}
The reduced model predictions of the observations can be written as
\begin{equation}
	\db_r(\ub, \Phi) = \BO \Phi \wb_r(\ub)
\end{equation}
leading to the simplification of equation \ref{eq:pred_grad},
\begin{equation}
	\nabla_{\ub} \dbp = \BO \Phi  \nabla_{\ub} \wb_r +  \nabla_{\ub} \epsn(\ub)
\end{equation}
$g(\wb_r, \ub) = 0$ everywhere implies that $\nabla_{\ub} g(\wb_r, \ub) = 0$. Expanding the total derivative yields,
\begin{equation}
	\nabla_{\ub} g(\wb_r, \ub) =  \dfrac{\partial g(\wb_r, \ub)}{\partial \wb_r} \nabla_{\ub} \wb_r + \dfrac{\partial g(\wb_r, \ub)}{\partial \ub} = 0
\end{equation} 
and rearranging to obtain the gradient of the reduced state with respect to the parameters gives,
\begin{equation}
 \nabla_{\ub} \wb_r = \left(\dfrac{\partial g(\wb_r, \ub)}{\partial \wb_r}\right)^{-1}\dfrac{\partial g(\wb_r, \ub)}{\partial \ub}
\end{equation}
Substituting in equation \ref{eq:J_grad} leads to,
\begin{equation}
\nabla_{\ub} J(\ub) = (\db_{\text{obs}} - \dbp )^T\left[ \BO \Phi \left(\dfrac{\partial g(\wb_r, \ub)}{\partial \wb_r}\right)^{-1}\dfrac{\partial g(\wb_r, \ub)}{\partial \ub} + \nabla_{\ub} \epsn(\ub) \right] + \gamma \nabla_{\ub} \reg{\ub}
\end{equation}
Identifying the adjoint variables $\lambda \in \mathbb{R}^n$ as, 
\begin{equation}
\lambda^T = (\db_{\text{obs}} - \dbp )^T \BO \Phi \left(\dfrac{\partial g(\wb_r, \ub)}{\partial \wb_r}\right)^{-1} 
\end{equation}
we solve the adjoint equation
\begin{equation}
\left(\dfrac{\partial g(\wb_r, \ub)}{\partial \wb_r}\right)^T \lambda = (\BO \Phi)^T(\db_{\text{obs}} - \dbp )
\end{equation}
to evaluate the gradient of the objective function with respect to the parameters
\begin{equation}
\nabla_{\ub} J(\ub) = \lambda^T\dfrac{\partial g(\wb_r, \ub)}{\partial \ub} + (\db_{\text{obs}} - \dbp )^T \nabla_{\ub} \epsn(\ub) + \gamma \nabla_{\ub} \reg{\ub}
\end{equation}

\end{document}